\documentclass[fleqn,usenatbib]{mnras}

\usepackage{newtxtext,newtxmath}

\usepackage[T1]{fontenc}

\DeclareRobustCommand{\VAN}[3]{#2}
\let\VANthebibliography\thebibliography
\def\thebibliography{\DeclareRobustCommand{\VAN}[3]{##3}\VANthebibliography}

\usepackage{graphicx}	
\usepackage{amsmath}	
\usepackage{threeparttable} 
\usepackage{booktabs,caption}
\usepackage{subcaption}

\title[Cosmic $\gamma$-ray background radiation from SFGs]{The diffuse extragalactic $\gamma$-ray background radiation: star-forming galaxies are not the dominant component}

\author[J. Chen and T. Totani]{
Junling Chen$^{1}$\thanks{E-mail: chenjl96@g.ecc.u-tokyo.ac.jp}
and Tomonori Totani$^{1, 2}$
\\
$^{1}$Department of Astronomy, School of Science, the University of Tokyo, Bunkyo-ku, Tokyo 113-0033, Japan\\
$^{2}$Research Center for the Early Universe, School of Science, the University of Tokyo, Bunkyo-ku, Tokyo 113-0033, Japan \\
}

\date{Accepted XXX. Received YYY; in original form ZZZ}

\pubyear{\the\year{}}

\begin{document}
\label{firstpage}
\pagerange{\pageref{firstpage}--\pageref{lastpage}}
\maketitle

\begin{abstract}
Star-forming galaxies (SFGs) are considered to be an important component of the diffuse extragalactic gamma-ray background (EGB) radiation observed in 0.1 -- 820 GeV, but their quantitative contribution has not yet been precisely determined. In this study, we aim to provide the currently most reliable estimate of the contribution of SFGs based on careful calibration with $\gamma$-ray luminosities of nearby galaxies and physical quantities (star formation rate, stellar mass, and size) of galaxies observed by high-redshift galaxy surveys. Our calculations are based on the latest database of particle collision cross-sections and energy spectra of secondary particles, and take into account not only hadronic but also leptonic processes with various radiation fields in a galaxy. We find that SFGs are not the dominant component of the unresolved EGB measured by \textit{Fermi}; the largest contribution is around 50\% -- 60\% in the 1 -- 10 GeV region, and the contribution falls rapidly in lower and higher energy ranges. This result appears to contradict a previous study, which claimed that SFGs are the dominant component of the unresolved EGB, and the origin of the discrepancy is examined. In calculations of cosmic-ray production, propagation, and interaction in a galaxy, we try models developed by two independent groups and find that they have little impact on EGB. 
\end{abstract}




\section{Introduction}
\label{section:intro}



The observation of the extragalactic gamma-ray background (EGB) began in the 1970s when the OSO-3 satellite first detected diffuse high-energy $\gamma$-ray radiation \citep{kraushaar1972high}, confirming its isotropic nature and extragalactic origin. Around the new century, COMPTEL \citep{COMPTEL} and EGRET \citep{EGRET} on the Compton Gamma-Ray Observatory measured the EGB intensity in a higher level of accuracy. These efforts laid the groundwork for the precise measurements achieved by the Fermi Gamma-Ray Space Telescope (\textit{Fermi}) nowadays. \textit{Fermi} has measured EGB in the energy range of 100 MeV--820 GeV \citep{ackermann2015spectrum}. \textit{Fermi} observations have shown that approximately half of the EGB has been resolved into detected individual sources, but the origin of the unresolved EGB continues to be debated \citep{fornasa2015nature}. Several candidate $\gamma$-ray sources have been proposed as the components of the unresolved EGB, including active galactic nuclei (AGNs), millisecond pulsars, dark matter annihilation, and in this work we consider the contribution from star-forming galaxies (SFGs).  

Cosmic ray (CR) particles, including protons and leptons, are accelerated by shocks in supernova remnants (SNRs) to high energy in SFGs. High energy CR protons will interact with the interstellar medium (ISM) within a galaxy by the proton-proton (pp) collisions, producing $\pi^0$ and $\pi^\pm$ which rapidly decay as $\pi^0\rightarrow2\gamma$, $\pi^+\rightarrow e^++\nu_e+\nu_\mu+\Bar{\nu}_\mu$, and $ \pi^-\rightarrow e^-+\Bar{\nu_e}+\nu_\mu+\bar{\nu}_\mu$. Therefore, the $\pi^0$ decay process will produce high energy $\gamma$-ray photons directly. High energy leptons, either directly accelerated by SNR shocks (primary) or produced by $\pi^\pm$ decay processes (secondary), can also produce high-energy $\gamma$-ray photons by bremsstrahlung and inverse Compton (IC) scattering. Generally, the differential cross-sections of both $\gamma$-ray photon and lepton production maximize when the energy of secondary particles is a few percents of the energy of primary CR protons. 


Most previous studies estimated the contribution of SFGs in the unresolved EGB at 10 -- 50\% (see Fig. A1 of \citealt{owen2022extragalactic}). Early studies were based on a simple positive correlation between star formation rates (SFRs) and $\gamma$-ray luminosities of galaxies, simply assuming that $\gamma$-ray spectra of galaxies are the same as those of nearby galaxies  \citep{fields2002,fields2010cosmic,makiya2011contribution} or the spectrum of $\pi^0$ decay process  \citep{stecker2011components}. Recent studies used more detailed physical models regarding the production, propagation, and interaction of CR particles  \citep{sudoh2018high, shimono2021prospects, roth2021diffuse, peretti2019, peretti2020contribution, owen2022extragalactic, ambrosone}. Among these, \cite{roth2021diffuse} (hereafter R21) estimated a substantially higher contribution from SFGs than many previous studies, which fully accounts for the unresolved EGB by SFGs alone, by estimating $\gamma$-ray luminosities of individual galaxies detected in a high-redshift galaxy survey \citep{CANDELS1, CANDELS2}. 

In this work, we present a new estimate of the SFG contribution to the unresolved EGB, based on our previous model constructed by \cite{sudoh2018high, shimono2021prospects} with updates and improvements on various aspects. In this model, $\gamma$-ray luminosity of a galaxy is calculated from its physical quantities (e.g. SFR, stellar mass and size), and the strength of this model is that its prediction agrees well with the observed $\gamma$-ray luminosities of nearby galaxies. \cite{sudoh2018high} estimates the SFG contribution to the unresolved EGB based on this model, but $\gamma$-ray luminosities of high-$z$ galaxies were calculated theoretically by a semi-analytic galaxy formation model, without direct use of observational data for high-$z$ galaxies. In this work we apply the approach of \cite{roth2021diffuse} and use the observed physical quantities of high-$z$ galaxies in the CANDELS GOODS-South sample \citep{CANDELS1, CANDELS2, CANDELS_data1, CANDELS_data2} to calculate $\gamma$-ray luminosities of individual galaxies. This provides a reliable estimate of the SFG contribution, consistent with observational data, both in $\gamma$-ray luminosities of nearby galaxies and in physical quantities of high-$z$ galaxies. For models that calculate CR production, propagation, and interaction to predict $\gamma$-ray luminosities of galaxies, we carefully consider the contribution of CR lepton emission, which was not considered in our previous works \citep{sudoh2018high,shimono2021prospects}. We also apply an independent model used in \cite{roth2021diffuse}, instead of our own, to test model dependence. These analyses are expected to shed new light on the SFG contribution to the unresolved EGB, which has varied widely in previous studies.

This paper is arranged as follows. In Section \ref{Methods}, we describe methods including our model for the $\gamma$-ray emission from SFGs, the processing of galaxy sample data, and the calculation of the cosmic background. Section \ref{Results} presents the main results, including the fitting of our model to nearby galaxies and the EGB flux and spectrum from SFGs. In Section \ref{Discussion}, we compare our results with some previous studies, and discuss potential EGB sources beyond SFG. Conclusions are presented in Section \ref{Conclusions}. Throughout this work, we assume a flat $\Lambda$CDM cosmology with $\Omega_{\mathrm M}=0.3$, $\Omega_\Lambda=0.7$ and $H_0=70 $ $\mathrm {km \, s^{-1}Mpc^{-1}}$.

\section{Methods}\label{Methods}
Here we present our model of $\gamma$-ray emission from a galaxy, and the application of the model to the CANDELS GOODS-S sample. Our model is based on a former model constructed by \cite{sudoh2018high} (hereafter S18), which estimates $\gamma$-ray emission from a galaxy by four properties of galaxy: stellar mass $M_*$, gas mass $M_{\rm gas}$, effective radius $R_{\mathrm{e}}$ and SFR $\psi$, with some improvement. 


\subsection{Model of CR proton emission}
CR protons accelerated by SNR shocks are the major contributor to the $\gamma$-ray emission from SFGs. The CR proton production rate (particle number per unit proton energy $E_{\mathrm p}$) can be related to SFR as
\begin{equation}
    \frac{dN_{\mathrm p}}{dtdE_{\mathrm p}} = C \left(\frac{\psi}{\mathrm M_{\odot} \rm yr^{-1}}\right)\left(\frac{E_{\mathrm p}}{\mathrm {GeV}}\right)^{-\Gamma_{\mathrm {inj}}} , 
    \label{eq1}
\end{equation}
where the energy spectral index $\Gamma_{\mathrm {inj}}$ is typically 2.2--2.4 for the Milky Way (MW) \citep{index1, index2}, which is determined by the observed spectrum of the diffuse Galactic background radiation. We assume $\Gamma_{\mathrm {inj}}=2.2$ for all galaxies throughout this work. The normalization factor $C$ can be theoretically calculated. 
We set the range of stellar mass as 0.1--150 $ M_\odot$, the mass threshold of core-collapse supernovae as 8 $M_\odot$, the energy injected into CR protons from one supernova event as $10^{50}$ erg, the Salpeter initial mass function (IMF) \citep{salpeter1955luminosity}, and the energy range of CR particle as $E_{\rm p} = m_{\rm p} c^2$--\,$4\times10^{15}$ eV ($m_{\rm p}$ is the proton mass and $c$ is the velocity of light), unless otherwise stated. It should be noted that $E_{\rm p}$ is the total proton energy including rest mass, but when the CR energy is integrated over $E_{\rm p}$ to be compared with the supernova energy, kinetic CR energy excluding rest mass is considered. 
The relation between SN rate and SFR can be expressed as 
\begin{equation}
    \mathrm{SN\ rate} = \frac{\psi\int_{8 M_\odot}^{150 M_\odot}\xi(m)dm}{\int_{0.1 M_\odot}^{150 M_\odot}m\xi(m)dm}. 
\end{equation}
Therefore, $C$ can be theoretically calculated as
\begin{equation}
    C = \frac{\int_{8 M_\odot}^{150 M_\odot}\xi(m)dm}{\int_{0.1 M_\odot}^{150 M_\odot}m\xi(m)dm}\frac{10^{50}\,\mathrm{erg}}{\int_{m_{\rm p} c^2}^{4\times10^{15} \mathrm{eV}}\left(\frac{E_{\mathrm p}-m_{\rm p} c^2}{\mathrm {10^9 eV}}\right)\left(\frac{E_{\mathrm p}}{\mathrm {10^9 eV}}\right)^{-\Gamma_{\mathrm {inj}}}\,dE_{\rm p}}
    \label{C}
\end{equation}
where $\xi(m)$ is Salpeter IMF. We omit the units of SFR in equation \ref{C}. 

In this case, $C = 3.69\times 10^{33} \mathrm {s^{-1}eV^{-1}}$. However, in this work we adopt $C = 3.83\times 10^{33} \mathrm {s^{-1}eV^{-1}}$, which is determined by fitting to the observed $\gamma$-ray luminosities of six nearby galaxies, as will be described in Section  \ref{fitting}.  

A fraction $f_{\mathrm {cal}}$ of CR protons will interact with interstellar medium (ISM) at most once before they escape from the source galaxy, and this can be expressed as $f_{\mathrm {cal}}=1-{\mathrm {exp}}(-t_{\mathrm {esc}}/t_{\mathrm {pp}})$, where $t_{\mathrm {esc}}$ is the escape timescale of a CR particle from the galaxy and $t_{\mathrm {pp}}$ is the interaction timescale with ISM by proton-proton (pp) collisions. The interaction timescale can be written as $t_{\mathrm {pp}}= (n_{\mathrm {gas} } \sigma_{\mathrm {pp}}c)^{-1}$, where $n_{\mathrm {gas}}$ is the proton number density of ISM, and the inelastic part of the total cross-section of pp collision $\sigma_{\mathrm {pp}}(E_{\mathrm p})$ is given by equation 79 in \cite{Kelner}. The number density is modeled as $n_{\mathrm {gas}}=M_{\mathrm {gas}}/(2\pi R_{\mathrm {eff}}^2H_{\mathrm g}m_{\mathrm p}\mu _{\mathrm p})$, where $\mu_{\mathrm p}=1.17$ is the number ratio of nucleons to a proton, and $H_{\mathrm g}$ the scale height of the gas disk of the galaxy. This model assumes $H_{\mathrm g}\propto R_{\mathrm {eff}}$ for all galaxies, which is consistent with nearby galaxy observations \citep{height}, and the ratio is determined by the values of MW: $H_{\mathrm g}=150$ pc \citep{Mo} and $R_{\mathrm {eff}}=6.0$ kpc \citep{r_mw}. 

We consider two mechanisms of CR proton escape from a galaxy: diffusion and advection, and therefore the escape timescale is determined as $t_{\mathrm {esc}}=min(t_{\mathrm {diff}},t_{\mathrm {adv}})$. These two timescales are estimated from galactic properties as $t_{\mathrm {diff}}=H^2_{\mathrm g}/[2D(E_{\mathrm p})]$ and $t_{\mathrm {adv}}=H_{\mathrm g}/\sigma$. Here $D(E_{\mathrm p})$ is the diffusion coefficient for galaxies, and $\sigma$ is the escape velocity from the gravitational potential of the galactic disk, which is determined from $H_{\mathrm g}$ and the surface density of total mass $\Sigma_{\mathrm {tot}} = (M_*+M_{\mathrm {gas}})/(\pi R_{\mathrm {eff}}^2)$ by the vertical structure of an isothermal sheet: $G \, \Sigma_{\mathrm {tot}}=\sigma^2/(2\pi H_{\mathrm g})$  \citep{Mo}. The diffusion coefficient is estimated in a standard manner based on the Larmor radius $R_{\mathrm L}=2.0\times 10^{-7}(E_{\mathrm p}/{\mathrm {GeV}})(B/6{\mu \mathrm G})^{-1}$pc and the coherence length $l_0={\mathrm {min}}(30{\mathrm {pc}},H_{\mathrm g})$ of the interstellar magnetic field assuming Kolmogorov-type turbulence. The diffusion coefficient is then expressed as 
\begin{equation}
    D(E_{\mathrm p})= \begin{cases}
    \frac{cl_0}{3}\left[ \left(\frac{R_{\mathrm L}}{l_0}\right)^{\frac{1}{3}}+\left(\frac{R_{\mathrm L}}{l_0}\right)^2 \right] & (R_{\mathrm L}\leq \sqrt{H_{\mathrm g}l_0}) \\
    \frac{cH_{\mathrm g}}{3} & (R_{\mathrm L}\geq \sqrt{H_{\mathrm g}l_0})
    \end{cases}
    \label{eq2}
\end{equation}
 \citep{D_coefficient}. The magnetic field strength of a galaxy is estimated by the assumption that the energy density of the magnetic field is close to that of supernova explosions injected into ISM on the advection timescale $t_{\mathrm {adv}}$: $B^2/(8\pi)=\eta \, E_{\mathrm {SN}} \, r_{\mathrm {SN}} \, t_{\mathrm {adv}} / (2\pi R_{\mathrm {eff}}^2H_{\mathrm g})$, where $E_{\mathrm {SN}} = 10^{51}$ erg and $r_{\mathrm {SN}}$ are the explosion energy released by one supernova and the SN rate, respectively. The value of $\eta = 0.23$ is set to reproduce the typical value of the Galactic magnetic field strength $B = 6 \, \mu \mathrm G$ \citep{B_mw}.  

In the end we calculate the $\gamma$-ray spectrum by CR proton interactions in a galaxy, which can be expressed as
\begin{equation}
    \left. \frac{dN_{\gamma}}{dtdE_{\gamma}}\right|_{\mathrm {proton}} = \int_{m_{\rm p}c^2}^\infty \left[\frac{1}{\sigma_{\mathrm {pp}}}\frac{d\sigma_\gamma}{dE_\gamma}\right]f_{\mathrm {cal}}\frac{dN_{\mathrm p}}{dtdE_{\mathrm p}}dE_{\mathrm p},
    \label{eq3}
\end{equation}
where $E_\gamma$ is the $\gamma$-ray photon energy, and $d\sigma_\gamma/dE_\gamma$ (a function of $E_{\rm p}$) is the differential cross-section of $\gamma$-ray photons produced by a $\pi^0$ decay. We use \texttt{AAfrag2.0} \citep{aafrag} code to calculate the differential cross-section.


\subsection{Model of CR lepton emission}\label{lep}
CR leptons (electrons and positrons) also contribute to the production of $\gamma$-ray in SFGs, especially in the lower energy band of $\gamma$-ray spectrum. CR electrons can be accelerated by SNR shocks similarly to CR protons (primary production), meanwhile both CR electrons and positrons appear in the decay of charged pions $\pi^{\pm}$ produced by pp collisions (secondary production). We assume that the spectral index of primary electrons is the same as protons, and hence the injection rate per unit electron energy is
\begin{equation}
    \left. \frac{dN_{\mathrm {e^-}}}{dtdE_{\mathrm {e^-}}}\right|_1 =  C_{\rm 1e} \left(\frac{\psi}{\mathrm {M_{\odot}yr^{-1}}}\right)\left(\frac{E_{\mathrm {e^-}}}{\mathrm {GeV}}\right)^{-\Gamma_{\mathrm {inj}}} \ ,
    \label{eq4}
\end{equation}
where the normalization factor $C_{\rm 1e}=0.012\,C$ is determined by
assuming that the injection energy of CR electrons is 
1.2\% of protons \citep{Strong} in the energy range of $10^8$ to $10^{11}$ eV. 

For secondary leptons, we adopt a similar manner to the $\gamma$-ray production of CR protons since they are all from pion decay. The injection rate spectrum of secondary leptons is then
\begin{equation}
    \left. \frac{dN_{\mathrm {e^\pm}}}{dtdE_{\mathrm {e^\pm}}}\right|_{2} = \int_{m_{\mathrm {p}} c^2}^\infty \left[\frac{1}{\sigma_{\mathrm {pp}}}\frac{d\sigma_{\mathrm {e^\pm}}(E_{\mathrm p})}{dE_{\mathrm {e^\pm}}}\right]f_{\mathrm {cal}}\frac{dN_{\mathrm p}}{dtdE_{\mathrm p}}dE_{\mathrm p},
    \label{eq5}
\end{equation}
where $d\sigma_{\mathrm {e^\pm}}/dE_{\mathrm {e^\pm}}$ is the differential cross-section of $e^+{\text { or }} e^- $ production by $\pi^\pm$ decay. We again use \texttt{AAfrag2.0} to calculate these differential cross-sections. Thus the total CR lepton injection spectrum is $dN_l/(dtdE_l)=dN_{\mathrm {e^-}}/(dtdE_{\mathrm {e^-}})|_1+dN_{\mathrm {e^+}}/(dtdE_{\mathrm {e^+}})|_2+dN_{\mathrm {e^-}}/(dtdE_{\mathrm {e^-}})|_2$. It should be noted that the only difference between the secondary electrons and positrons in the injection spectra is their differential cross-section $d\sigma_{\mathrm {e^\pm}}/dE_{\mathrm {e^\pm}}$. In the energy loss processes in ISM which will be described below, all the primary electrons and the secondary $e^\pm$ particles are treated as the same particles with the spectrum $dN_l/(dtdE_l)$. 

Unlike the case of CR protons, CR leptons may experience multiple energy loss processes before escaping the galaxy. CR leptons lose their energy by four major processes: collisional ionization, synchrotron radiation, bremsstrahlung, and inverse Compton scattering. The formulae of energy loss rates of these processes are taken from literature: \cite{Schlickeiser} for ionization and bremsstrahlung, \cite{synchrontron} for synchrotron,  and \cite{IC_losstime} for inverse Compton scattering. These can be written as follows:
\begin{equation}
    \left. \frac{dE_l}{dt}\right|_{\mathrm {ion}} = \frac{9}{4}\sigma_{\mathrm T}c\,m_{\mathrm e}c^2n_{\mathrm {gas}}\left[\ln\gamma + \frac{2}{3}\ln\left(\frac{m_{\mathrm e}c^2}{15{\mathrm {eV}}} \right) \right]
    \label{eq6}
\end{equation}

\begin{equation}
    \left. \frac{dE_l}{dt}\right|_{\mathrm {brems}} = \frac{3}{\pi}\alpha\,\sigma_{\mathrm T}c\,m_{\mathrm e}c^2\gamma\,n_{\mathrm {gas}}\begin{cases}
        \ln\gamma+\ln 2-\frac{1}{3}&\gamma\lesssim 15 \\
        \frac{1}{8}\Phi_{\mathrm {1,H}}\left(\frac{1}{4\alpha\gamma}\right)&\gamma\gtrsim 15
    \end{cases}
    \label{eq8}
\end{equation}

\begin{equation}
    \left. \frac{dE_l}{dt}\right|_{\mathrm {sync}} = \frac{1}{6\pi}\sigma_{\mathrm T}c\,B^2\gamma^2\beta^2
    \label{eq7}
\end{equation}

\begin{equation}
    \left. \frac{dE_l}{dt}\right|_{{\mathrm {IC,}}i} = \frac{20}{\pi^4}\sigma_{\mathrm T}c\,\gamma^2u_{{\mathrm {rad,}}i}Y(\Gamma_i).
    \label{IC_loss}
\end{equation}
In these equations, $\sigma_{\mathrm T}$ is the Thomson cross-section, $m_{\mathrm e}c^2$ is the rest energy of lepton, $\gamma=E_l/(m_{\rm e} c^2)$ is the lepton Lorentz factor, $\beta$ is the CR velocity normalized by $c$, $\alpha$ is the fine structure constant, and $\Phi_{\mathrm {1,H}}$ and $Y$ are dimensionless numerical fitting function, which can be found in the corresponding references. The argument of the function $Y$ is $\Gamma_i=4\gamma E_{{\mathrm {peak,}}i}/(m_{\rm e} c^2)$, where $i$ corresponds to different radiation fields in the galaxy: starlight, dust emission and cosmic microwave background (CMB). The treatments of the energy density $u_{{\mathrm {rad,}}i}$ and the spectral peak photon energy $E_{{\mathrm {peak,}}i}$ of each radiation field will be described in Section  \ref{section:CANDELS}. The energy loss timescale for the $j$-th energy loss process among the four mentioned above is defined as
\begin{equation}
    t_{{\mathrm {loss,}}j} = E_l \left( \left. \frac{dE_l}{dt}\right|_j \right)^{-1} , 
    \label{loss_time}
\end{equation}
while the total energy loss timescale is $t_{\mathrm {loss}} = (\sum_j t^{-1}_{{\mathrm {loss,}}j})^{-1}$. The number of electrons existing in a galaxy at a given time is determined by the shorter of the energy loss ($t_{\rm loss}$) or escape ($t_{\rm esc}$) time scales, and this can be approximated as
\begin{equation}
    \frac{dN_l}{dE_l}=\frac{dN_l}{dtdE_l}(t_{\mathrm {loss}}^{-1}+t_{\mathrm {esc}}^{-1})^{-1}\ ,
    \label{El_density}
\end{equation}
where $t_{\mathrm {esc}}$ is calculated by the same manner as CR proton case. 

Among the four energy loss processes, bremsstrahlung and inverse Compton scattering can produce $\gamma$-ray photons. We calculate the $\gamma$-ray spectrum from bremsstrahlung  and inverse Compton scattering by the following formulae:
\begin{equation}
    \left.\frac{dN_\gamma}{dtdE_\gamma}\right|_{\mathrm {brems}} = \frac{cn_{\mathrm {gas}}}{E_\gamma}\int_{E_\gamma}^\infty\sigma_{\mathrm {BS}}(E_\gamma,E_l)\frac{dN_l}{dE_l} dE_l \ , 
    \label{brems_gamma_spectrum}
\end{equation}
\begin{equation}
    \left.\frac{dN_\gamma}{dtdE_\gamma}\right|_{{\mathrm {IC,}}i} = \frac{3}{4}\sigma_{\mathrm T}c\frac{u_{{\mathrm {rad,}}i}}{E_{{\mathrm {peak,}}i}^2}\int_{E_{l, \mathrm {min}}}^{\infty}\frac{G(a,\Gamma_i)}{\gamma^2}\frac{dN_l}{dE_l} dE_l \ ,
    \label{IC_gamma_spectrum}
\end{equation}
where $\sigma_{\mathrm {BS}}$ is the differential cross-section for nonthermal bremsstrahlung \citep{Schlickeiser, brems1, brems2_IC}, 
\begin{equation}
G(a,\Gamma_i)=2a{\ln}a+(1+2a)(1-a)+\frac{(\Gamma_i a)^2(1-a)}{2(1+\Gamma_i a)}
\end{equation}
and $a=E_\gamma/\Gamma_i(E_l-E_\gamma)$ \citep{peretti2019, brems2_IC}.   
The total CR lepton $\gamma$-ray emission spectrum is then simply obtained as:
\begin{equation}
    \left.\frac{dN_\gamma}{dtdE_\gamma}\right|_{\mathrm {lepton}} = \left.\frac{dN_\gamma}{dtdE_\gamma}\right|_{\mathrm {brems}}+\left.\sum_i\frac{dN_\gamma}{dtdE_\gamma}\right|_{{\mathrm {IC,}}i} .
    \label{eq13}
\end{equation}

\subsection{Propagation of $\gamma$-ray photons and cascade effect}
$\gamma$-ray photons produced in SFGs will experience attenuation by infrared (IR) radiation within the galaxy. Still, this attenuation has little effect in the energy band that we are interested in (below TeV) \citep{owen2021}, and hence is ignored. After the $\gamma$-ray photons escape from the source galaxy, they are attenuated in the intergalactic space by extragalactic background light (EBL) photons. The primary $\gamma$-ray flux at the Earth from a galaxy at redshift $z$ can be calculated as
\begin{equation}
    \frac{dF_\gamma}{dE_\gamma}=\frac{(1+z)^2}{4\pi d_{\mathrm L}^2}\frac{dN_\gamma[E_\gamma(1+z)]}{dtdE_\gamma}
    \, \exp(-\tau_{\mathrm {EBL}}),
    \label{flux}
\end{equation}
where $d_{\mathrm L}$ is the luminosity distance of the galaxy, $dN_\gamma[E_\gamma(1+z)]/dtdE_\gamma$ is the $\gamma$-ray photon luminosity from a galaxy per unit rest-frame photon energy at energy $E_\gamma(1+z)$, and $\tau_{\mathrm {EBL}}$ is the optical depth from attenuation by EBL photons. We adopt $\tau_{\mathrm {EBL}}$ data from the model of \cite{inoue_EBL}. 

Meanwhile, the attenuated $\gamma$-ray photons interacting with EBL photons produce $e^\pm$ via pair-production. The secondary $e^\pm$ pairs interact with CMB photons via IC scattering subsequently. We build a simple model to describe the cascade effect of $\gamma$-ray flux from each galaxy based on the model of \cite{Inoue_cascade}. Our model bases on some assumptions: 1) the cascade effects all happen near the source galaxies; 2) the energy of secondary $e^\pm$ pair $\epsilon_{\mathrm e}\approx 0.5E_\gamma$; 3) reabsorption of cascade photons is negligible. The pair-production rate can be expressed as
\begin{equation}
    \frac{dN_{\mathrm e}}{dtd\epsilon_{\mathrm e}} = \frac{dN_\gamma[E_\gamma(1+z)]}{dtdE_\gamma}\frac{dE_\gamma}{d\epsilon_{\rm e}}[1-\exp(-\tau_{\mathrm {EBL}})] \ .
\end{equation}
According to equation \ref{IC_loss},\ref{loss_time},\ref{El_density}, 
the pair-production spectrum is 
\begin{equation}
    \frac{dN_{\mathrm e}}{d\epsilon_{\mathrm e}} = \frac{dN_{\mathrm e}}{dtd\epsilon_{\mathrm e}}t_{\mathrm {loss,IC}} \ ,
\end{equation}
and following equation \ref{IC_gamma_spectrum}, the cascade spectrum can be expressed as
\begin{equation}
     \left.\frac{dN_\gamma}{dtdE_\gamma}\right|_{\mathrm {cascade}} = \frac{3}{4}\sigma_{\mathrm T}c\frac{u_{\mathrm {rad,CMB}}}{E_{\mathrm {peak,CMB}}^2}\int_{\epsilon_{ \mathrm {e,min}}}^{\infty}\frac{G(a,\Gamma_i)}{\gamma^2}\frac{dN_{\mathrm e}}{d\epsilon_{\mathrm e}}\ d\epsilon_{\mathrm e} \ .
\end{equation}
The cascade flux at the Earth from a galaxy at redshift $z$ can be calculated the same as equation \ref{flux}.

\subsection{Application to the galaxy sample}
In order to calculate EGB flux based on our model, physical quantities such as SFRs of galaxies at high redshifts are required. We use the galaxy sample constructed by R21, which is based on CANDELS (Cosmic Assembly Near-infrared Deep Extragalactic Legacy Survey) \citep{CANDELS1, CANDELS2, CANDELS_data1} GOODS-S (Great Observatories Origins Deep Survey-South) \citep{CANDELS_data2}. CANDELS is one of the largest surveys conducted by the Hubble Space Telescope, and GOODS-S is one of the most widely used deep fields with comprehensive multi-wavelength data, covering an area of 173.00 arcmin$^2$. The original galaxy catalog of CANDELS GOODS-S includes 34,930 galaxies, and R21 selected 22,278 galaxies after excluding those with uncertain parameters for some reasons (e.g. having bright AGNs). The R21 catalog provides SFRs, stellar masses, effective radii, and redshifts, but we need more data to estimate radiation fields which are necessary for the modeling of IC emission. When we compare the R21 sample to the latest version of the CANDELS GOODS-S catalog \citep{Rainbow2, Rainbow1} in the Rainbow database, \footnote{US: \href{http://arcoiris.ucolick.org/Rainbow_navigator_public}{http://arcoiris.ucolick.org/Rainbow\_navigator\_public/}, \\ and Europe: \href{http://rainbowx.fis.ucm.es/Rainbow_navigator_public/}{http://rainbowx.fis.ucm.es/Rainbow\_navigator\_public/}} a small fraction of galaxies in the former are not found in the latter. Therefore we use 22,087 galaxies that are included in both the R21 and CANDELS GOODS-S catalogs, and physical quantities of galaxies are taken from the catalog of \cite{Rainbow2, Rainbow1}.

\subsubsection{IMF correction}
\label{section:IMF}
In this work the normalization factor $C$ in equation \ref{eq1} is determined by fitting to the observation of nearby galaxies. As the physical quantities of galaxies estimated from observations depend on IMF, it is important to ensure that there are no discrepancies in IMFs assumed in the nearby galaxy data used for the $C$ calibration and the high-redshift CANDELS data. For the nearby galaxies, we adopt the observational data collected by \cite{shimono2021prospects} which calculate SFRs and stellar masses assuming the Salpeter IMF, which is a simple power-law function across all stellar mass range. On the other hand, the CANDELS sample is based on the Chabrier IMF \citep{chabrier}, which is described by a power-law form at $\gtrsim 1M_\odot$ and a lognormal form at $\lesssim 1M_\odot$. Therefore we convert SFRs and stellar masses of CANDELS into those for the Salpeter IMF by dividing by constant factors of 0.63 and 0.61, respectively \citep{Madau}.

\subsubsection{Data processing of the CANDELS sample}
\label{section:CANDELS}
SFRs, stellar masses, and photometric redshifts are directly extracted from the catalog of \cite{Rainbow2, Rainbow1}. In addition, we calculated other galactic properties including effective radii (half-light radii), gas masses, and radiation field energy densities using quantities in the original data. We calculate the effective radius at the rest-frame wavelength of 5000 {\AA} from those in the observed bands of F125W and F160W, using equation 2 in \cite{radius_correct}. Gas mass is calculated from SFRs and sizes using the Kennicutt-Schmidt law \citep{KS_law_pri, KS_law_98, KS_law_corr},
\begin{equation}
    \Sigma_{\mathrm {SFR}} = 10^{-10.1}\left(\frac{\Sigma_{\mathrm {gas}}}{M_\odot\mathrm {pc^{-2}}}\right)^{1.38}M_\odot{\mathrm 
 {yr^{-1} pc^{-2}}},
    \label{eq15}
\end{equation}
where the definition of surface density is similar to that for $\Sigma_{\mathrm {tot}}$. It should be noted that this formula has been corrected for the Salpeter IMF case. 

As mentioned in Section \ref{lep}, we consider three different radiation fields: starlight, dust emission, and CMB as the target photon field of the inverse Compton scattering emission. For CMB we set $u_{\mathrm {rad,CMB}}=4\sigma_{\mathrm {SB}}(1+z)^4T_0^4/c$ and $E_{\mathrm {peak,CMB}}=2.82k_{\mathrm B}(1+z)T_0$, where $\sigma_{\mathrm {SB}}$ is the Stefan–Boltzmann constant, $T_0=2.736\mathrm \ K$ and $k_{\mathrm B}$ is the Boltzmann constant. The radiation fields of starlight and dust emission depend on the physical properties of galaxies. The dust emission is mostly in IR bands, and the radiation energy density is estimated by IR luminosity $L_{\mathrm {IR}}$, as $u_{\mathrm {rad,dust}}=L_{\mathrm {IR}}/(c\pi R_{\mathrm {eff}}^2)$. In the CANDELS GOODS-S catalog, SFRs of galaxies that are detected in IR are calculated by the sum of SFRs estimated by IR and ultraviolet (UV) luminosities, i.e. SFR = SFR(IR) + SFR(UV). We then estimate IR luminosity $L_{\mathrm {IR}}$ of these galaxies from SFR(IR) given in the catalog using the SFR-$L_{\mathrm {IR}}$ relation of \cite{KS_law_98}. SFRs of IR-undetected galaxies are estimated by UV luminosity after correction about extinction (SFR$_{\rm corr}$). We therefore estimate IR luminosities of these galaxies from  $\rm SFR_{corr}-SFR(UV)$ and the SFR-$L_{\mathrm {IR}}$ relation. The energy density of starlight is estimated using luminosities in UV, optical and near-IR bands ($L_{\mathrm {opt}}$) as $u_{\mathrm {rad,star}}=L_{\mathrm {opt}}/(c\pi R_{\mathrm {eff}}^2)$. We estimate $L_{\mathrm {opt}}$ using the bolometric luminosity $L_{\mathrm {bol}}$ in the range of 92 \AA\;to 160 $\mu$m given in the catalog, as $L_{\mathrm {opt}} = L_{\mathrm {bol}}-L_{\mathrm {IR}}$. Finally, we set the peak energy of starlight to be 5000 \AA\ in the rest-frame, and that of dust emission to be $E_{\mathrm {peak,dust}}=2.82k_{\mathrm B}T_{\mathrm {dust}}$, where $T_{\mathrm {dust}}$ is estimated from the relation of \cite{dust_tem}: $T_{\mathrm {dust}}=98(1+z)^{-0.065}+6.9\log(\mathrm {SFR/M_\odot yr^{-1}})$.

\subsection{Integrating to the cosmic background flux}
After calculating the $\gamma$-ray fluxes from all 22,087 galaxies, we can integrate the result into the EGB flux from SFGs. However, the galaxies detected by CANDELS do not cover all galaxies in the universe, and contributions from galaxies below the detection limit must also be considered. The CANDELS field is only a small region of the entire sky and may be affected by local increases or decreases in galaxy number density along the line of sight, due to large-scale structure. We correct for these effects by comparing the cosmic SFR with SFRs of CANDELS galaxies at each redshift bin, assuming that the properties of galaxies below the detection limit are not significantly different from those of CANDELS galaxies. \cite{Madau} gives the best-fitting function of cosmic SFR density $\psi$ with the assumption of the Salpeter IMF:
\begin{equation}
    \psi(z) = 0.015\frac{(1+z)^{2.7}}{1+[(1+z)/2.9]^{5.6}} M_\odot\mathrm {yr^{-1}Mpc^{-3}} \ ,
    \label{eq16}
\end{equation}
which is obtained by integrating observed luminosity functions below their detection limits. We divide the whole redshift range $z = $ 0 -- 10 into 100 bins of size $\Delta z=0.1$. The cosmic SFR within the GOODS-S area in the $j$-th redshift bin is ${\mathrm {SFR}}_{{\mathrm {cosmic}},j}=\int_{z_j}^{z_{j+1}}\psi (dV/dz) dz$, where $dV/dz$ is the comoving volume per unit redshift in the survey area, and the total CANDELS SFR is ${\mathrm {SFR}}_{{\mathrm {CANDELS}},j}=\sum_{i(j)}{\mathrm {SFR}}_{i}$, where $i$ is for the sum of all the CANDELS galaxies in the $j$-th redshift bin. We then calculate EGB from SFGs as:
\begin{equation}
    \Phi(E_\gamma) = \sum_j f_{{\rm corr}, j} \sum_{i(j)} \left(\frac{dF_\gamma}{dE_\gamma}\right)_{i} \Delta \Omega^{-1}
    \label{eq:Phi}
\end{equation}
where $\Delta \Omega$ is the survey area of CANDELS, and
\begin{equation}
    f_{{\rm corr}, j}  = \frac{{\mathrm {SFR}}_{{\mathrm {cosmic}},j}}{{\mathrm {SFR}}_{{\mathrm {CANDELS}},j}}  .
    \label{eq:f_corr}
\end{equation}

\section{Results}\label{Results}

\subsection{Fitting result to nearby galaxies} \label{fitting}

Here we check whether the predictions of our $\gamma$-ray emission model are consistent with the observed $\gamma$-ray luminosities of nearby galaxies, and determine the value of the normalization parameter $C$ for the background radiation flux calculation.
We follow the fitting method and the nearby $\gamma$-ray galaxy sample provided by \cite{shimono2021prospects}, and the observed galaxy parameters are summarized in Table.\ref{tab:nearby_galaxies}. The SFRs and stellar masses in this sample are all based on the Salpeter IMF. 

\begin{figure}
    \centering
    \begin{subfigure}[b]{0.45\textwidth}
        \centering
        \includegraphics[width=\textwidth]{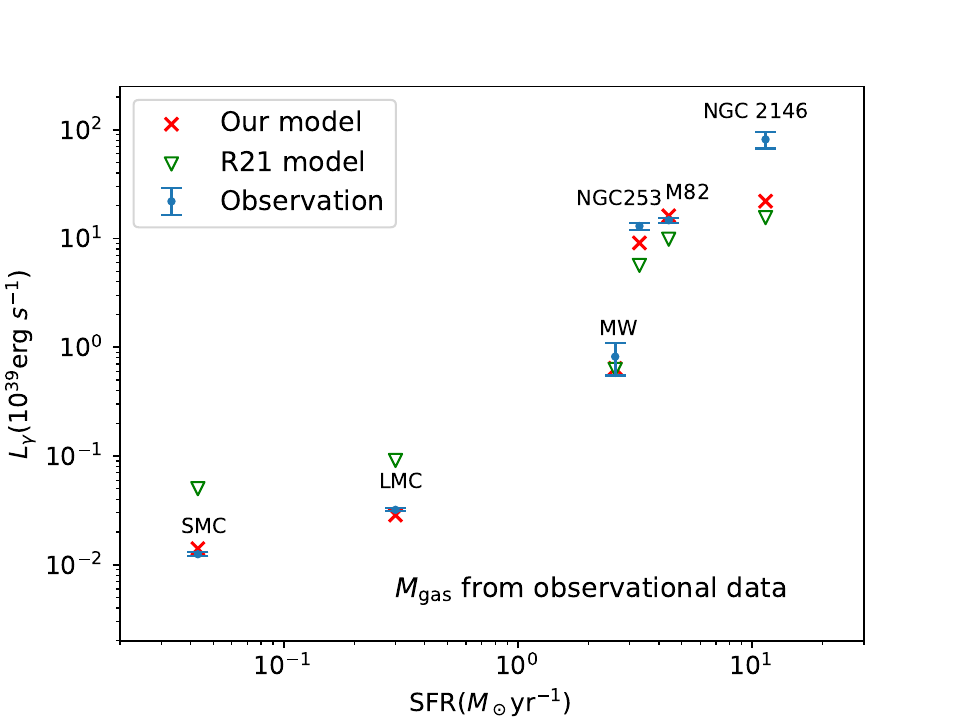}
        \label{fig:panelA}
    \end{subfigure}
    \hfill
    \begin{subfigure}[b]{0.45\textwidth}
        \centering
        \includegraphics[width=\textwidth]{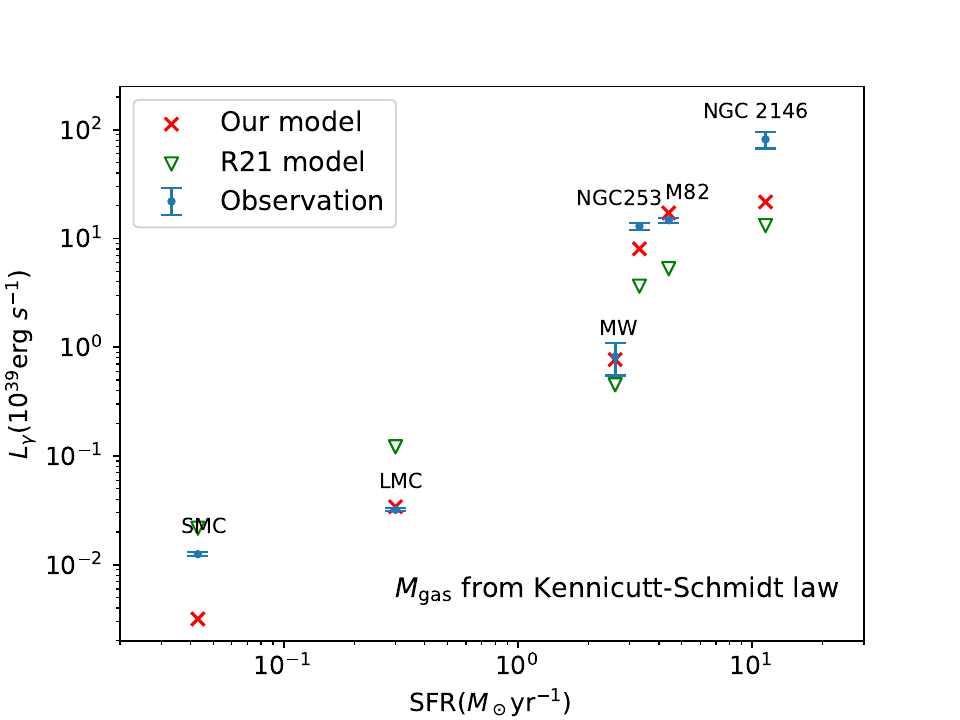}
        \label{fig:panelB}
    \end{subfigure}
    \caption{$\gamma$-ray luminosities of nearby galaxies predicted by our model and R21 model, in comparison with the observed luminosities. Upper panel: directly observed gas masses are used for $M_{\mathrm {gas}}$. Lower panel: $M_{\mathrm {gas}}$ is calculated by the Kennicutt-Schimidt law from SFR and size. The $\gamma$-ray energy range is 0.1 -- 100 GeV for MW, and 0.1 -- 800 GeV for all others. }
    \label{fig:nearby_gals}
\end{figure}

The upper panel of Fig. \ref{fig:nearby_gals} compares the predictions of $\gamma$-ray luminosities by our model using the full set of observed quantities: SFR, $M_{\rm gas}$, $M_{\rm star}$, and $R_{\rm eff}$ (and also $L_{\rm IR}$ and $L_{\rm opt}$ for the lepton components). Here, the normalization parameter of $C = 3.83\times 10^{33} \mathrm {s^{-1}eV^{-1}}$ is used, which is obtained by the $\chi^2$ fit to the data in this figure. The luminosities predicted by our model are in good agreement with the observed values except for NGC 2146, and the value of the normalization factor $C$ is also in close agreement with the original theoretical value ($3.69 \times 10^{33}$). The discrepancy of NGC 2146 may indicate a limitation of our model, which assumes a simple disk geometry and a uniform matter distribution within. This assumption is likely too simple, especially in some starburst galaxies, whose morphology and matter distribution are quite complex. Instead, it may suggest a contribution other than star-forming activity to the $\gamma$-ray luminosity of this galaxy. 
\cite{NGC2146_AGN} recently found that the center of NGC 2146 may be a low-luminosity AGN.
It should also be noted that the distance to this galaxy is considerably further than other galaxies, and hence the observed galaxy parameters may have larger uncertainties.

The lower panel of Fig. \ref{fig:nearby_gals} is the same as the upper panel, but with only SFR, $M_{\rm star}$, $R_{\rm eff}$, $L_{\rm IR}$ and $L_{\rm opt}$ as inputs, as in the high-$z$ CANDELS data, while $M_{\rm gas}$ is calculated using the Kennicutt-Schmidt law for our model and the extended Schmidt law for R21, following their method. The agreement between the theoretical and observed values has worsened for SMC, but otherwise, there is little change.

For comparison, the same calculations are performed using an independent model of $\gamma$-ray emission from SFGs used in R21. The R21 model assumes the Chabrier IMF, so the input galaxy parameters are corrected about this. In addition to the galaxy parameters listed in Table \ref{tab:nearby_galaxies}, the R21 model also requires the disk scale height $h$, which is calculated according to R21, using equations relating it to other galaxy parameters. Here, the normalization parameter $\phi$ in the R21 model (corresponding to $C$ in our model but by slightly different definition) is set to $5.44 \times 10^{33} \ \rm s^{-1}eV^{-1} M_\odot^{-1} yr$, assuming the Chabrier IMF in the mass range 0.1--50 $M_\odot$. This value is different from that adopted by R21 ($7.15 \times 10^{33}$) for the same IMF assumption, but we use our own value because we could not reproduce the R21's value by our calculation (see also Section \ref{sec:prev_work} later). The $\phi$ value depends on the integration range over proton energy, and the value we adopted ($5.44 \times 10^{33}$) is the case for the integration down to proton momentum $p_{\rm p} = 1$ GeV/$c$. The $\phi$ value changes to $4.8 \times 10^{33}$ and $5.8 \times 10^{33}$ when the integration is down to $p_{\rm p} = 0$ or $p_{\rm p}$ corresponding to $E_{{\rm p}, k} = m_{\rm p} c^2$, respectively, where $E_{{\rm p},k}$ is the kinetic proton energy excluding the rest mass. 

The results for the R21 model are also shown in Fig. 
\ref{fig:nearby_gals}. 
Overall, the R21 model shows a similar degree of agreement with the observed data as our model. We also test the model dependence by calculating the EGB flux from SFGs using this model, which will be presented in Section \ref{sec:prev_work}. 

We additionally calculate the spectra of nearby galaxies, as presented in Appendix \ref{other_spectra}. For galaxies except MW, the observational data are obtained from the most recent \textit{Fermi}-LAT Fourth Source Catalog Data Release 4 (4FGL-DR4) \citep{4FGL1,4FGL}\footnote{\href{https://fermi.gsfc.nasa.gov/ssc/data/access/lat/14yr_catalog/}{https://fermi.gsfc.nasa.gov/ssc/data/access/lat/14yr\_catalog/}}. Since the $\gamma$-ray luminosity and spectrum of MW cannot directly be observed, we compare our result to the observed and model spectra of the Galactic diffuse emission reported in \cite{index1}, and find a reasonable agreement in the spectral shape. It is noteworthy that the spectra of individual galaxies vary significantly due to the unique environment in each galaxy.

\begin{table*}
	\centering
	\caption{Physical properties of GeV-detected galaxies used to determine normalization factor $C$}
	\label{tab:nearby_galaxies}
        \begin{threeparttable}

	\begin{tabular}{lcccccccc} 
		\hline
		Objects & $L_\gamma\mathrm{(0.1-800\,GeV)^a}$ & ${\mathrm {SFR^b}}$ & $M_{\mathrm {gas}}^{\mathrm c}$ & $M_{\mathrm {star}}^{\mathrm d}$ & $R_{\mathrm {eff}}^{\mathrm e}$ & $D^{\mathrm{f}}$& $L_{\mathrm {IR}}^{\mathrm g}$ & $L_{\mathrm {optical}}^{\mathrm h}$ \\
               & ($10^{39}{\mathrm {erg s}}^{-1}$) & ($M_\odot \mathrm {yr^{-1}}$) & ($10^9M_\odot$) & ($10^9M_\odot$) & (kpc) & (Mpc) & ($10^{43}{\mathrm {erg s}}^{-1}$) & ($10^{43}{\mathrm {erg s}}^{-1}$) \\
		\hline
		MW & $0.82\pm0.27$ & 2.6 & 4.9 & 64.3 & 6.0 & -- & 5.2 & 19.6\\
		LMC & $0.032\pm0.001$ & 0.3 & 0.59 & 1.8 & 2.2 & 0.05 & 0.253 & 0.509\\
		SMC & $0.0125\pm0.0005$ & 0.043 & 0.46 & 0.3 & 0.7 & 0.06 & 0.0277 & 0.153\\
        NGC 253 & $13\pm1$ & 3.3 & 3.2 & 54.4 & 0.5 & 3.5 & 10.5 & 7.46\\
        M82 & $14.7\pm0.7$ & 4.4 & 4.7 & 21.9 & 0.3 & 3.3 & 22.5 & 3.07\\
        NGC 2146 & $81.4\pm14.2$ & 11.4 & 10.4 & 87.1 & 1.7 & 17.2 & 45.0 & 7.27\\
		\hline
	\end{tabular}
        \begin{tablenotes}
            \item[a] $\gamma$-ray luminosities taken from \cite{nearby_luminosity} except for MW from \cite{nearby_luminosity_mw1}. Energy range is 0.1 -- 100 GeV for MW but 0.1 -- 800 GeV for all other galaxies. 
            \item[b] Star-formation rates calculated from \cite{nearby_sfr_mw2, makiya2011contribution} for MW, and \cite{nearby_sfr1, nearby_sfr2, nearby_sfr3} for all other galaxies. 
            \item[c] Total gas masses (atomic and molecular hydrogen) calculated from \cite{gas_mw1, gas_mw2_ngc2146} for MW; \cite{gas_lmc1, gas_lmc2_smc, gas_lmc3_smc} for LMC; \cite{gas_smc1, gas_lmc2_smc, gas_lmc3_smc} for SMC; \cite{gas_ngc2531, gas_ngc2532, gas_ngc2533} for NGC 253; \cite{gas_M821, gas_M822, gas_M823} for M82; and \cite{gas_mw2_ngc2146, gas_ngc21462} for NGC 2146. 
            \item[d] Stellar mass taken from \cite{star_mw} for MW, and \cite{star_other,star_other2} for other galaxies.
            \item[e] Effective radii taken from \cite{r_mw} for MW, \cite{r_lmc} for LMC, \cite{r_smc} for SMC, \cite{r_NGC253_M82} for NGC 253 and M82, and \cite{r_ngc2146} for NGC 2146.
            \item[f] Distances taken from \cite{nearby_sfr2} for LMC, SMC and NGC 2146, \cite{distance_ngc253} for NGC 253, and \cite{distance_m82} for M82. 
            \item[g] Total IR luminosities taken from \cite{nearby_luminosity} except for MW from \cite{Strong}. 
            \item[h] Optical luminosities taken from \cite{Strong} for MW, \cite{optical_LMC_SMC} for LMC and SMC, \cite{optical_ngc253} for NGC 253, \cite{optical_m82} for M82, and \cite{optical_ngc2146} for NGC 2146. 
            \item[] See \cite{shimono2021prospects} for more details of the properties in the table. 
        \end{tablenotes}
        \end{threeparttable}
\end{table*}



\subsection{Cosmic $\gamma$-ray spectrum from SFGs}\label{background result}
\begin{figure}
	\includegraphics[width=\columnwidth]{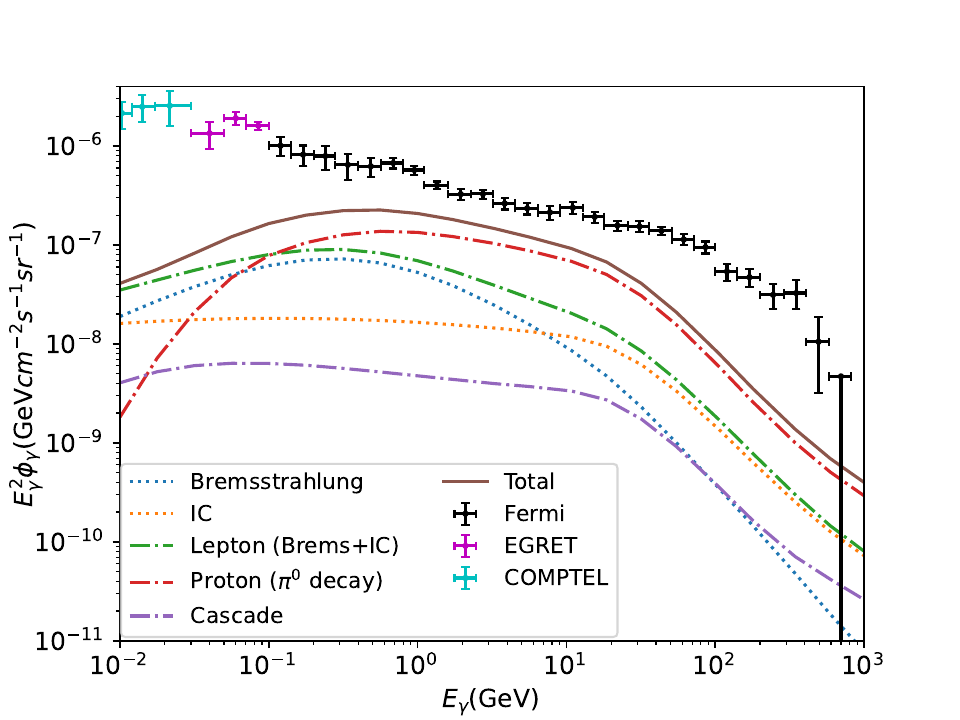}
    \caption{The EGB spectrum from SFGs predicted by our model, in comparison with the unresolved EGB data. Different components are shown as: bremsstrahlung (blue dotted), IC scattering (orange dotted), total lepton emission (green dash-dotted), proton emission (i.e. $\pi^0$ decay, red dash-dotted), cascade effect (purple dash-dotted), and total emission (brown solid). \textit{Fermi}-LAT data (black points) are from \protect\cite{ackermann2015spectrum}, EGRET data (magenta points) from \protect\cite{EGRET}, and COMPTEL data (cyan points) from \protect\cite{COMPTEL}.}
    \label{fig:background}
\end{figure}

Fig. \ref{fig:background} presents the contribution to the EGB from SFGs by our model. SFGs can explain a significant fraction of the unresolved EGB flux measured by \textit{Fermi}-LAT in the range of 1-10 GeV, but are severely deficient below 1 GeV or above 10 GeV. 

In this figure we also show the contribution of different processes that produce $\gamma$-ray photons in SFGs. In most of the energy band, emission from CR protons ($\pi^0$-decay) dominates the spectrum. But in the lower energy band, emission from CR leptons is comparable to and even higher than the emission from CR protons. Since the $\pi^0$ mass is 135 MeV, the $\pi^0$-decay gamma-ray flux decreases rapidly below 0.1 GeV. The predicted spectrum softens above 10 GeV, mainly because of the attenuation by interaction with EBL photons.


\section{Discussion}\label{Discussion}
\subsection{Comparison to previous works}
\label{sec:prev_work}
Our model is based on the model constructed by \cite{sudoh2018high} and \cite{shimono2021prospects}, but improves in several ways. We employ \texttt{AAfrag2.0} to compute the differential cross-section in a pp collision. The reasons we choose \texttt{AAfrag2.0} instead of \cite{Kelner} or some other previous works are: 1) \texttt{AAfrag2.0} is a well-tested new model that covers a larger energy range, providing detailed spectra of secondary particles (photons, electrons, positrons, etc.); 2) \texttt{AAfrag2.0} offers the differential cross-section data directly, prominently saving computing time and resources.
We establish a comprehensive model of CR lepton emission, especially for the IC scattering process: we consider the radiation fields of starlight, dust emission and CMB separately. S18 uses a semi-analytical model of cosmology galaxy formation by \cite{mitaka}, called Mitaka model. In this work, we adopt a real galaxy sample, CANDELS GOODS-S, combined with a best-fit function of the cosmic star formation history.

\begin{figure}
	\includegraphics[width=\columnwidth]{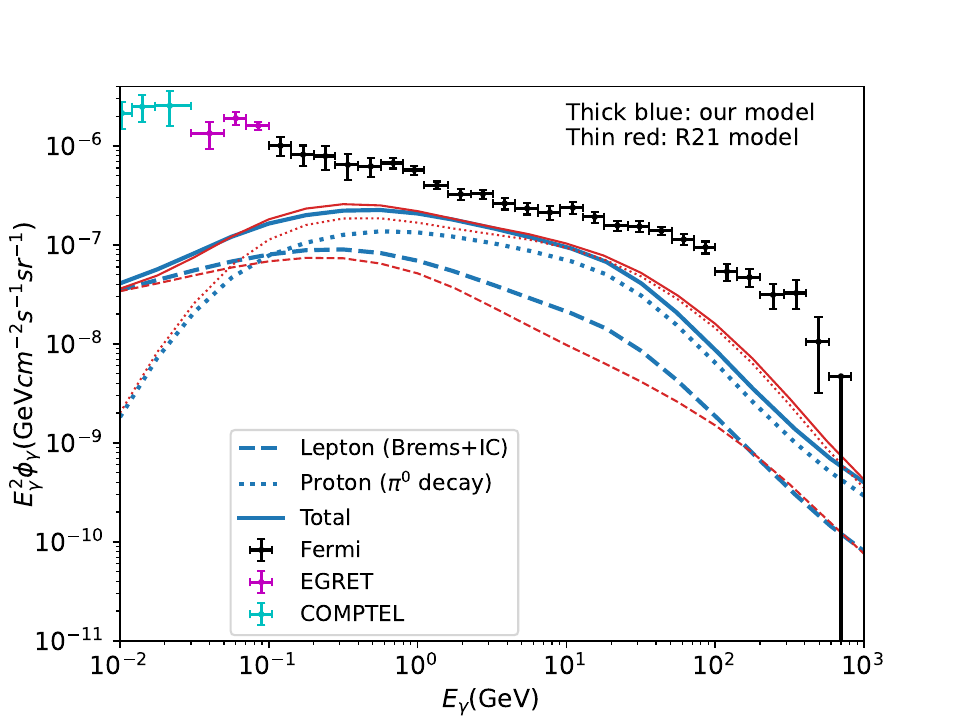}
    \caption{The EGB spectrum from SFGs predicted by our model (blue thick curves), in comparison with our another calculation using the R21 modeling framework (red thin curves). 
    }
    \label{fig:sudoh_v_roth}
\end{figure}

\begin{figure}
	\includegraphics[width=\columnwidth]{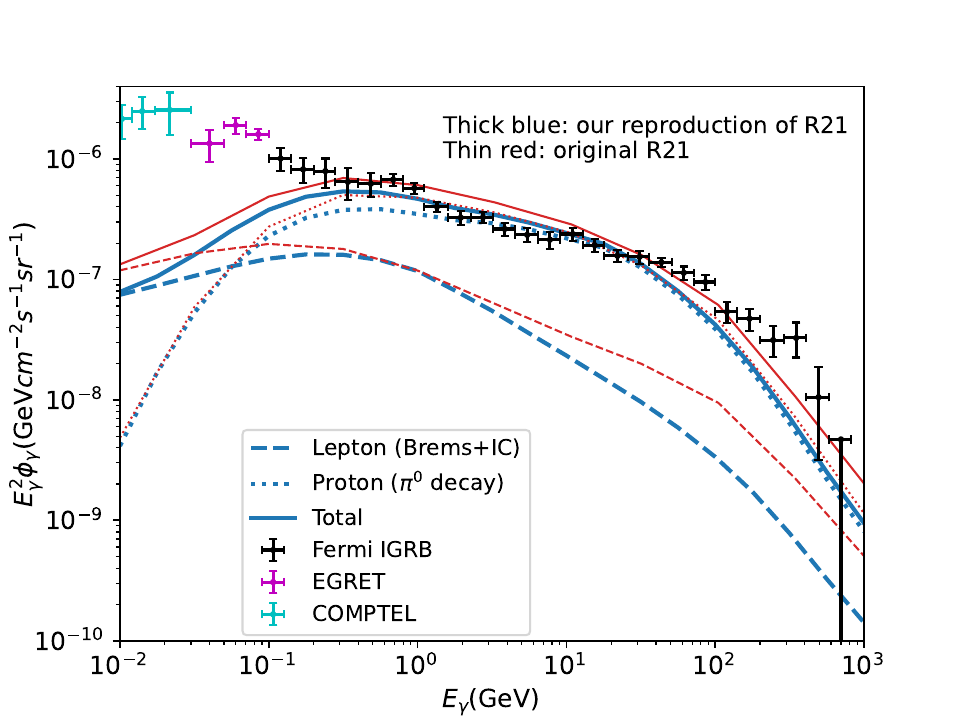}
    \caption{  The EGB spectrum from SFGs by our calculation, intentionally using inconsistent IMFs: the Chabrier IMF for input galaxy parameters of CANDELS while the Salpeter IMF for the cosmic SFR (red curves). For comparison, blue curves show the EGB spectrum from SFGs reported by R21. }
    \label{fig:roth_paper}
\end{figure}

As mentioned in Section \ref{section:intro},
\cite{roth2021diffuse} found a substantially higher EGB flux from SFGs than most previous models, which is sufficient to account for the entire unresolved EGB. To examine the origin of the discrepancy, we make another EGB flux calculation using the R21 modeling framework for $\gamma$-ray emission efficiency from SFGs. All input quantities are converted into those for the Salpeter IMF, as in our original model, and hence we use the normalization factor 
$\phi = 3.45 \times 10^{33} \ \rm s^{-1} eV^{-1} M_\odot^{-1} yr$ for this IMF, rather than $5.44 \times 10^{33}$ for the Chabrier IMF used in Section \ref{fitting}. Another difference from the original R21 modeling is the treatment of secondary lepton injection, for which we use our own modeling to save computation time. Fig. \ref{fig:sudoh_v_roth} compares the EGB flux by this computation to our own model prediction. These two calculations show a notable agreement, indicating that the EGB flux from SFGs is not much changed by the two independent modelings about $\gamma$-ray emission efficiency (e.g., production, propagation, and interaction of CR particles in a galaxy). 

We then try to reproduce the original R21 result in the following way. R21 adopts the normalization factor as $\phi = 7.15 \times 10^{33} \ \rm s^{-1} eV^{-1} M_\odot^{-1} yr$ assuming the Chabrier IMF. We could not reproduce this value, which should be $5.44 \times 10^{33}$ by our own calculation (see Section \ref{fitting}), but we adopt this R21 value here. 
However, we found that our calculated flux is still significantly lower than R21. There may be still other reasons for this discrepancy, including some miscalculations. Though we do not know the exact calculation procedures of R21, here we show a generally possible case. The cosmic SFR correction factor $f_{\rm corr}$ (equation \ref{eq:f_corr}) is calculated using cosmic SFR of \cite{Madau} that assumes the Salpeter IMF. Since the CANDELS SFRs are assuming the Chabrier IMF, one must correct this IMF difference for a reasonable calculation, but it is not clear whether this correction was made in R21. When we calculate without this IMF correction as a case of intentionally incorrect calculation, we find that the model EGB flux is increased by a factor of 1.6 (see Section \ref{section:IMF}), resulting in a reasonable agreement with that reported by R21 (Fig. \ref{fig:roth_paper}), especially for the $\pi^0$ decay component. Combined with the discrepancy about the normalization $\phi$, we believe that the EGB flux reported by R21 is overestimated by a factor of $1.6 \times 7.15 / 5.44 \sim 2.1$ compared to the correct value, for some unknown reasons. We cannot exclude the case where some miscalculations exist on our side, but our model flux aligns more closely with the results of most previous studies. 

In our EGB model, the ratio of the lepton to proton component is relatively higher than reported in some previous works \citep{peretti2020contribution, roth2021diffuse}. We identify two potential reasons for this discrepancy. First, there are differences in the pp collision models used: while previous works rely on the model by \cite{Kelner} to calculate the differential cross-section of secondary leptons from pp collisions, we employ \texttt{AAfrag2.0}, which produces higher cross-section values than those derived by \cite{Kelner}, especially for positron. Second, the assumed models for galactic magnetic fields differ. In this study, we assume that the energy density of the magnetic field is comparable to that of SN explosion injected into ISM on the advection scale. In contrast, other works, such as \cite{owen2022extragalactic}, adopt a turbulent dynamo model driven by SN events. These differences make it reasonable that our model predicts a higher CR lepton emission compared to previous studies.

\subsection{Gap between fluxes from SFGs and IGRB flux}
As shown in Section \ref{background result}, SFGs can explain a significant fraction ($\sim$50--60\%) of the unresolved EGB in the energy band of 1--10 GeV, but are insufficient in lower and higher energy ranges. Here we discuss some possible scenarios that account for this deficiency. In the energy band of $E_\gamma\lesssim$ 1 GeV, the major contributors may be some other extragalactic sources. The data of COMPTEL \citep{COMPTEL} and EGRET \citep{EGRET} imply some populations peaking around 10--100 MeV and extending up to 10 GeV by a softer power-law than SFGs. Millisecond pulsars \citep{siegal2011anisotropies} and some types of AGNs may be such populations. Some previous works investigated possible contributions from different populations of AGNs in detail: flat spectrum radio quasars (FSRQs) \citep{ajello2012luminosity, toda2020cosmological, linden} and misaligned AGNs (non-blazar AGNs) \citep{di2013diffuse, inoue2019high}.  In the energy band of $E_\gamma\gtrsim$ 10 GeV, the situation is more puzzling. Since all $\gamma$-ray photons with such high energy from distant extragalactic sources experience severe attenuation by EBL photons, we may need to consider relatively nearby sources in this energy range. BL Lac objects are one of the potential candidates, since they have a harder $\gamma$-ray spectrum than other AGNs, and relatively low luminosity among the blazer class (FSRQ and BL Lacs), potentially leaving nearby sources still undetected \citep{inoue2009blazar,di2014diffuse}. It is also possible that even more exotic populations are contributing in these energy ranges, e.g. dark matter annihilation in extragalactic galaxy halos and/or the MW halo \citep[e.g. ][]{Fermi_2015}.



\section{Conclusions}\label{Conclusions}
In this work, we constructed a new theoretical model of the EGB flux from cosmic-ray interactions in SFGs, based on the work of \cite{sudoh2018high} and \cite{shimono2021prospects}. Our model calculates $\gamma$-ray spectrum of a galaxy from some of its physical properties, including stellar mass, size, SFR, and radiation field strengths, by modeling production, propagation, and interactions of CR protons and leptons. The latest model of CR interaction cross-sections is used, and various components of radiation fields are considered to predict the inverse-Compton scattering emission. The model is calibrated by the observed $\gamma$-ray luminosities of six nearby galaxies. We then applied the model to the CANDELS GOODS-S sample to estimate the contribution of SFGs to the unresolved EGB. We consider this model to be the most reliable calculation to date, as it matches the $\gamma$-ray luminosity of nearby galaxies and is based directly on high-redshift galaxy survey observations.

According to our calculations, SFGs are an important component of the unresolved EGB at 1-10 GeV, but cannot account for all of it, only about 50-60\%. In the lower ($<$ 1 GeV) or higher ($>$ 10 GeV) energy range, the SFG contribution is even smaller in the observed unresolved EGB, falling below 10\% depending on photon energy. These results are broadly similar to previous estimates, except for R21, which claimed that SFGs are the dominant component of the unresolved EGB. We have examined this discrepancy by also performing calculations based on the framework of the R21 theoretical model. We found that the discrepancy is not caused by differences or uncertainties of the $\gamma$-ray emission modelings from SFGs, but that the EGB flux reported by R21 is likely to be overestimated by a factor of about 2.

Our results show that in the low- ($<$ 1 GeV) and high-energy ($>$ 10 GeV) regimes, astronomical objects other than SFGs are necessary to explain the unresolved EGB, giving some implications for future EGB studies. Various source populations may contribute to EGB in these regimes, including various types of AGNs and millisecond pulsars. One potential difficulty in building EGB theoretical models with these populations is that SFGs already have a significant contribution in the 1--10 GeV range, and the unresolved EGB flux must not be exceeded in this region. In the region above 10 GeV, $\gamma$-rays from extragalactic sources are subject to intergalactic absorption, and it is necessary to consider source populations of relatively small distances. The possibility of more exotic sources contributing, such as $\gamma$-rays from dark matter annihilation, should also be borne in mind.

\section*{Acknowledgements}

We are very grateful to Takahiro Sudoh and Ellis R. Owen for their helpful suggestions and stimulating discussions.
JC was supported by supported by JST SPRING, Grant Number JPMJSP2108. TT was supported by the JSPS/MEXT KAKENHI Grant Number 18K03692. 
This work has made use of the Rainbow Cosmological Surveys Database, which is operated by the Centro de Astrobiología (CAB/INTA), partnered with the University of California Observatories at Santa Cruz (UCO/Lick,UCSC). This research has also made use of the NASA/IPAC Extragalactic Database (NED), which is funded by the National Aeronautics and Space Administration and operated by the California Institute of Technology.

\section*{Data Availability}

The data underlying this article will be shared on reasonable request to the corresponding author.
 



\bibliographystyle{mnras}
\bibliography{example} 




\appendix

\section{Spectra of nearby galaxies}\label{other_spectra}

In this section, we show the $\gamma$-ray spectra of nearby galaxies mentioned in Section \ref{fitting}. For galaxies other than MW, we plot the observed spectra from 4FGL-DR4 \citep{4FGL1,4FGL} in addition. 

\begin{figure}
	\includegraphics[width=\columnwidth]{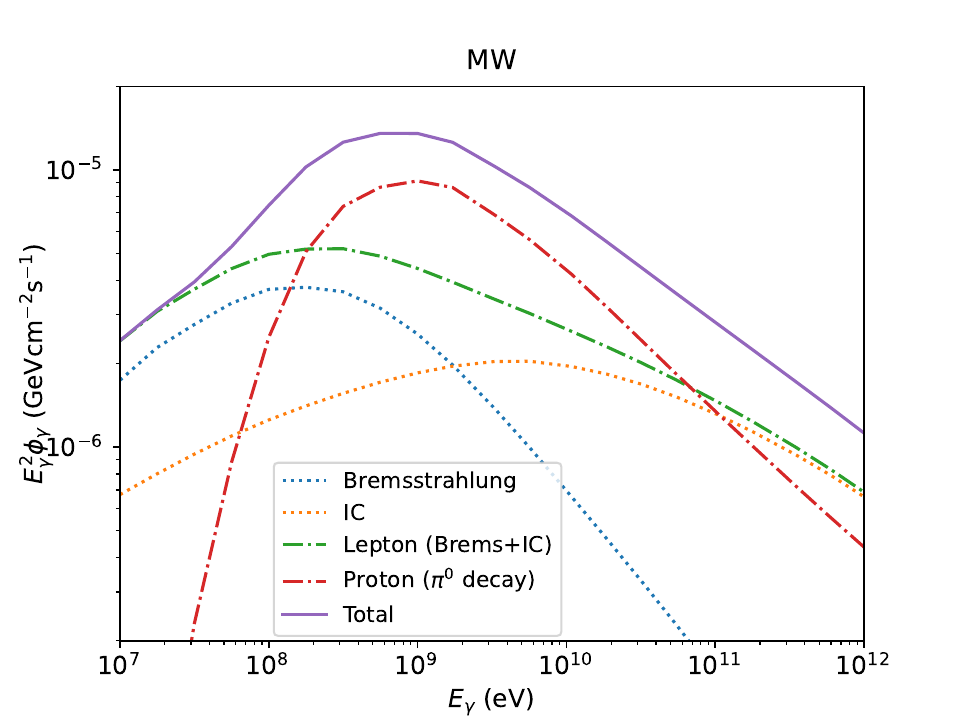}
    \caption{The $\gamma$-ray spectrum of MW predicted by our model, measured 8.3 kpc away from Galactic center (the distance from the sun to Galactic center). The line style is the same as in Fig.\ref{fig:background}}
    \label{fig:MW}
\end{figure}

\begin{figure}
    \centering
	\includegraphics[width=\columnwidth]{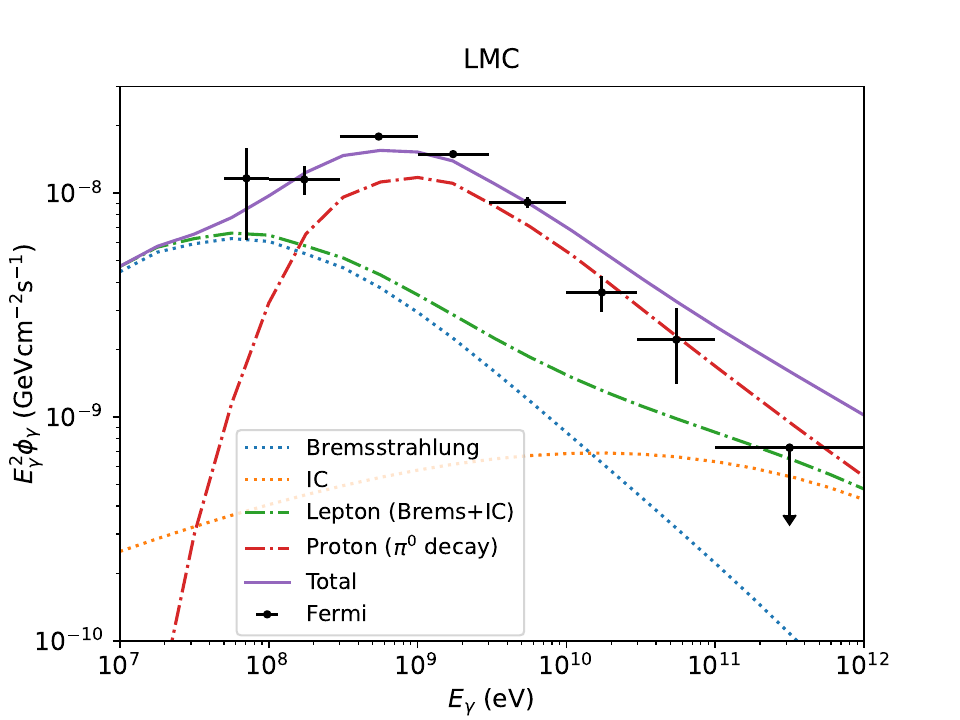}
    \caption{The same as Fig. \ref{fig:MW}, but for LMC and with observational data points from 4FGL-DR4. }
    \label{fig:LMC}
\end{figure}

\begin{figure}
    \centering
	\includegraphics[width=\columnwidth]{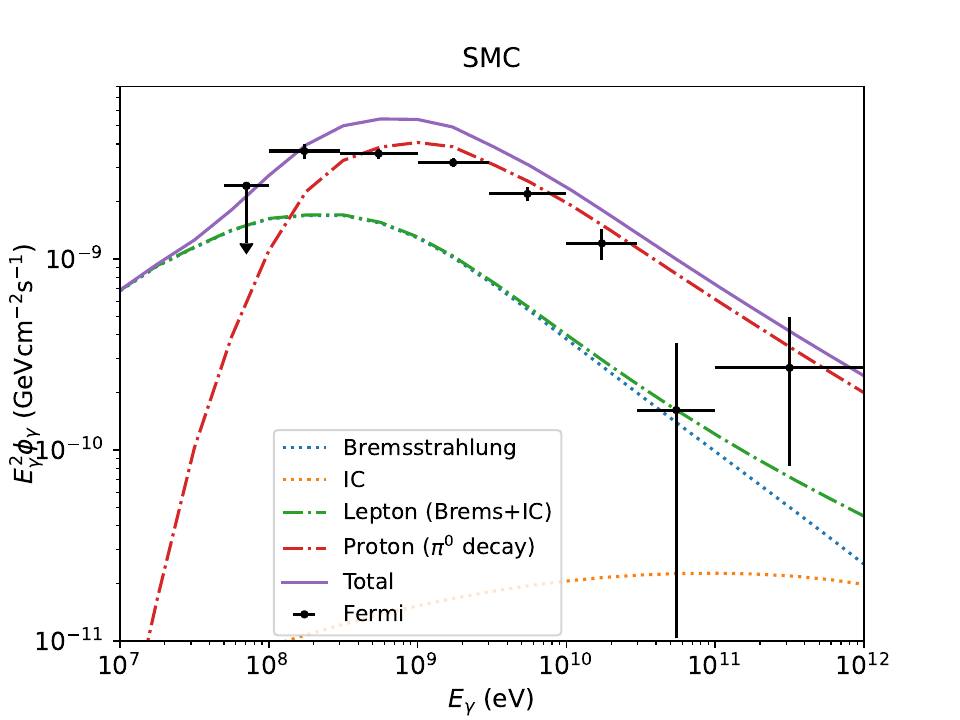}
    \caption{The same as Fig. \ref{fig:LMC}, but for SMC. }
    \label{fig:SMC}
\end{figure}

\begin{figure}
	\includegraphics[width=\columnwidth]{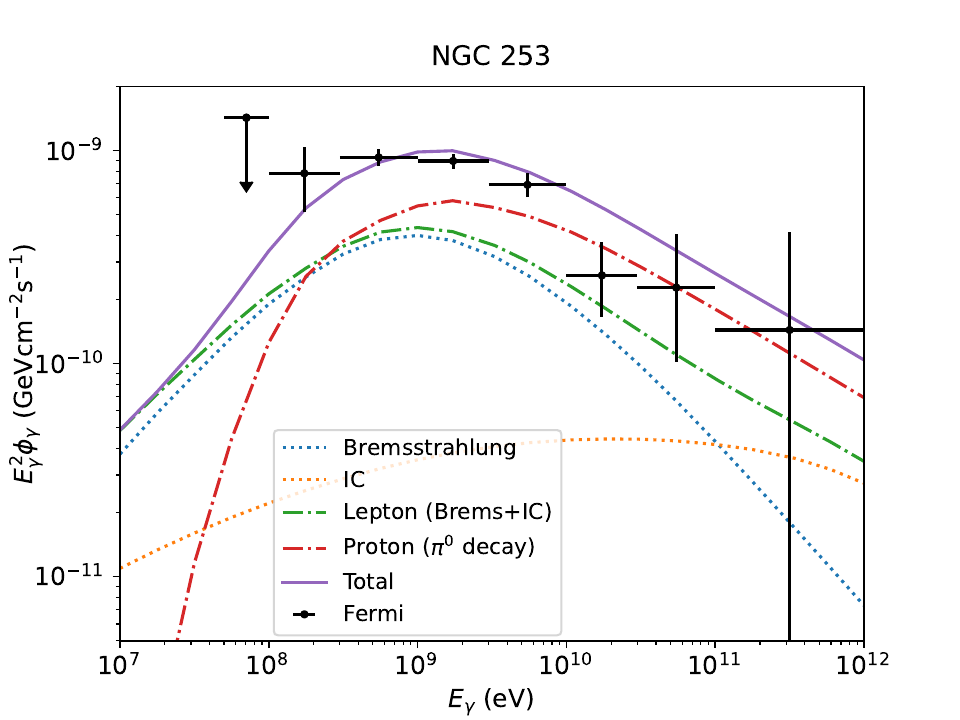}
    \caption{The same as Fig. \ref{fig:LMC}, but for NGC 253. }
    \label{fig:NGC253}
\end{figure}

\begin{figure}
    \centering
	\includegraphics[width=\columnwidth]{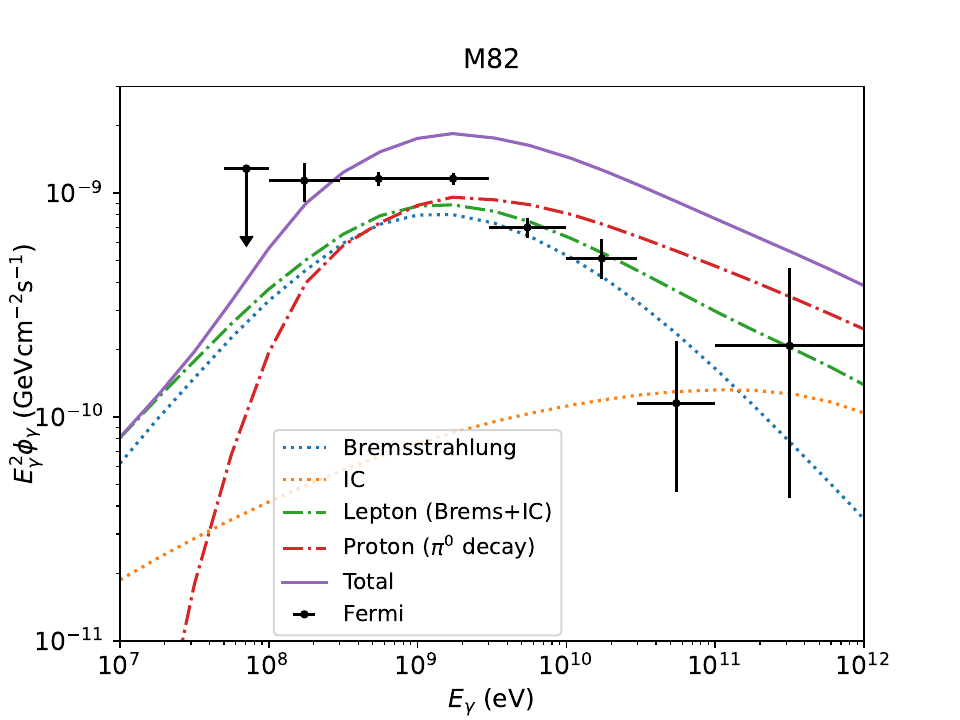}
    \caption{The same as Fig. \ref{fig:LMC}, but for M82. }    
    \label{fig:M82}
\end{figure}

\begin{figure}
    \centering
	\includegraphics[width=\columnwidth]{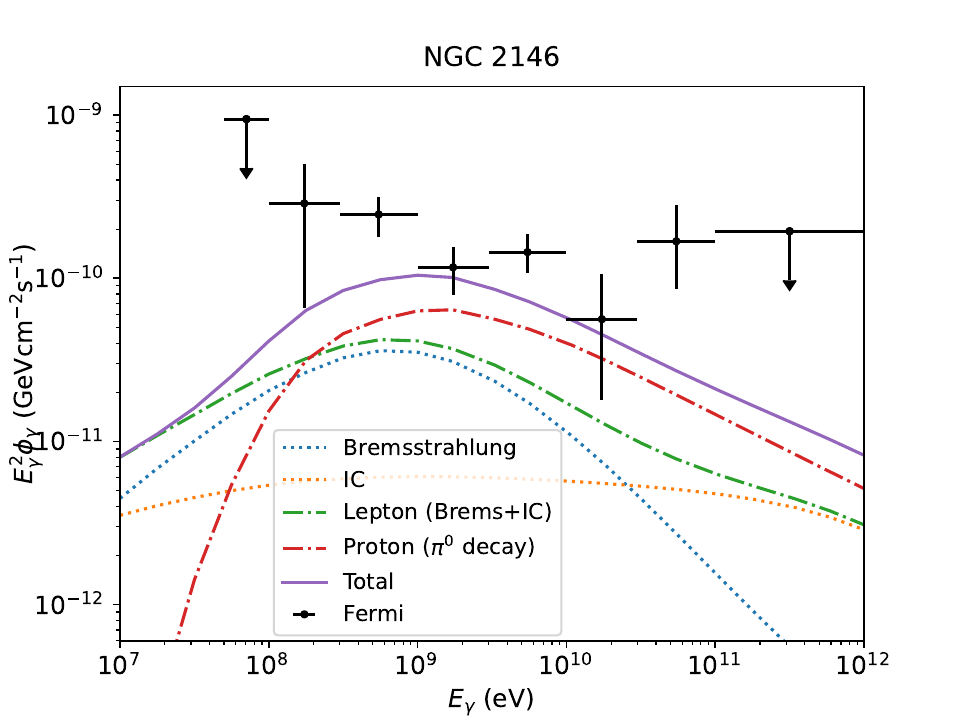}
    \caption{The same as Fig. \ref{fig:LMC}, but for NGC 2146. }  
    \label{fig:NGC2146}
\end{figure}



\bsp	
\label{lastpage}
\end{document}